# A New Screening Method for COVID-19 based on Ocular Feature Recognition by Machine Learning Tools

Yanwei Fu[1], Feng Li[3], Wenxuan Wang[2], Haicheng Tang[3],
Xuelin Qian[2], Mengwei Gu[1,4], Xiangyang Xue[1,2]

1, School of Data Science, Fudan University, 200433, China

2, School of Computer Science, Fudan University, Shanghai, 200433, China

3, Shanghai Public Health Clinical Center, Fudan University, Department of Critical Care, Shanghai, 201514, China

4, AIMOMICS (BVI) Holdings Limited

## Abstract

The Coronavirus disease 2019 (COVID-19) has affected several million people. With the outbreak of the epidemic, many researchers are devoting themselves to the COVID-19 screening system. The standard practices for rapid risk screening of COVID-19 are the CT imaging or RT-PCR (real-time polymerase chain reaction). However, these methods demand professional efforts of the acquisition of CT images and saliva samples, a certain amount of waiting time, and most importantly prohibitive examination fee in some countries. Recently, some literatures have shown that the COVID-19 patients usually accompanied by ocular manifestations consistent with the conjunctivitis, including conjunctival hyperemia, chemosis, epiphora, or increased secretions. After more than four months study, we found that the confirmed cases of COVID-19 present the consistent ocular pathological symbols; and we propose a new screening method of analyzing the eye-region images, captured by common CCD and CMOS cameras, could reliably make a rapid risk screening of COVID-19 with very high accuracy. We believe a system implementing such an algorithm should assist the triage management or the clinical diagnosis. To further evaluate our algorithm and approved by the Ethics Committee of Shanghai public health clinic center of Fudan University, we conduct a study of analyzing the eye-region images of 303 patients (104 COVID-19, 131 pulmonary, and 68 ocular patients), as well as 136 healthy people.  Remarkably, our results of COVID-19 patients in testing set consistently present similar ocular pathological symbols; and very high testing results have been achieved in terms of sensitivity and specificity.  We hope this study can be inspiring and helpful for encouraging more researches in this topic.

## Introduction

The Coronavirus disease 2019 (COVID-19) has affected several million people. The most important screening test for COVID-19 is to measure the body temperature. Such a measure, however, is very unreliable. As in [1], only 43.8% of COVID-19 patients on admission have the fever manifestation. Especially,  body temperature is a very inefficient screening method unable to handle the rising number of asymptomatic COVID-19 patients.

Notably, many COVID-19-positive patients were reported eye-related symptoms [2,3,6,7], such as conjunctival congestion, secretion, sagging eyelids, etc. Unfortunately, most ophthalmologists have not yet realized the recognition of ocular signs and symptoms STRONGLY associated with COVID-19. Particularly, we discussed this idea with many ophthalmologists. Our colleagues in ophthalmology do not have the deep understanding about the recognition of ocular signs and symptoms associated with COVID-19. On the other hand, it is totally understandable, since our colleagues in ophthalmology normally analyzed the patients by raw eyes, rather than the machine learning tools. Such thing always happened. For example, in the early COVID-19 cases, no doctor expected nor observed that there was loss of smell or taste to the confirmed COVID-19 patients.

Characterizing ocular manifestations in this cohort, should, in principle, serve as a new way for rapidly screening diagnostic test for COVID-19.Our conclusion in this paper is summarized as,

(1)The ocular symptoms are strongly associated with COVID-19;

(2)Build upon the recognition of ocular symptoms, we could utilize a new screening tool for the diagnostic test.

Until now, there is no report of utilizing neural network, or deep learning for the rapid screening method from eye-region photos. However, neural networks have indeed presented the good ability of learning representations, and extracted the sparse structures from complex, multi-dimensional, blurred, noise, and small sample data, such as CT-images. Such ability is beyond the capability of experts.

From April 1 to May 30, 303 patients admitted to Shanghai Public Health Clinical Center (SPHCC), Fudan University and AIMOMICS LAB were enrolled in this project, which included 104 patients with COVID-19, 131 with pulmonary diseases other than COVID-19, and 68 with ophthalmic diseases. In addition, 136 healthy volunteers were taken part in the research. Each person was taken five photos of eye regions using the common CCD and CMOS cameras, assisted by the doctor or health-care worker. The COVID-19 patients were confirmed by the RT-PCR detection for viral nucleic acids according to the seventh version guideline published by the National Health Commission of China. Among them, 3.85% COVID-19 patients were asymptomatic. Demographics and clinical information of patients with COVID-19 are shown in Table 1. The patients with pulmonary diseases other than COVID-19 were diagnosed as pulmonary fungal infection, bronchopneumonia, chronic obstructive pulmonary diseases and lung cancer, etc. The patients with ophthalmic diseases were diagnosed as trachoma, pinkeye, conjunctivitis, glaucoma, cataract and keratitis, etc. The healthy volunteers were collected from individuals who had taken physical examination and no obviously abnormal results were demonstrated. All the subjects were tested for the COVID-19 and the ones who were excluded COVID-19 did not show positive result for coronavirus during the following days. All the participants were aware of this experiment.  And the informed consents were obtained.

The photos were captured in a good lighting condition. Then the eye-regions in images were manually divided into several sector grids and labeled (Figure A).  The neural network tools were utilized to extract, quantify and concatenate the ocular features from each eye sector grid [4]; high-dimensional statistical algorithm was employed to recognize the ocular characteristics of each disease [5]. The 80% persons of each group were employed to train the model; the rest 20% people were mixed up to evaluate the model performance. The results show the AUC of each group is more than 0.95 (Figure B). This indicates the possibility of screening COVID-19 from ocular features; and the larger scale experiments with more patients will be conducted. The convenient method of eye-region image risk screening can help disease control researchers to fully understand the prevalence and pathogenicity of the virus in different age, time, region, climate, environment, occupation and population with basic diseases, and guide effective prevention and control measures against COVID-19.

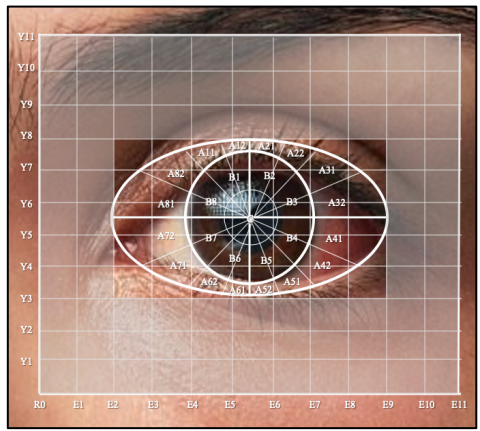

Figure A: Ocular Feature Extraction. In our study, we took the eye-region photos of confirmed patients, and divided the regions into fine-sections for analysis. The tools of neural network and high-dimensional statistics are utilized to extract, and quantify the features of sections from eye-regions, and further infer the disease. By virtue of this way, the machine learning tools can observe and recognize the subtle ocular signs and symptoms that are hard to be identified by ophthalmologists.

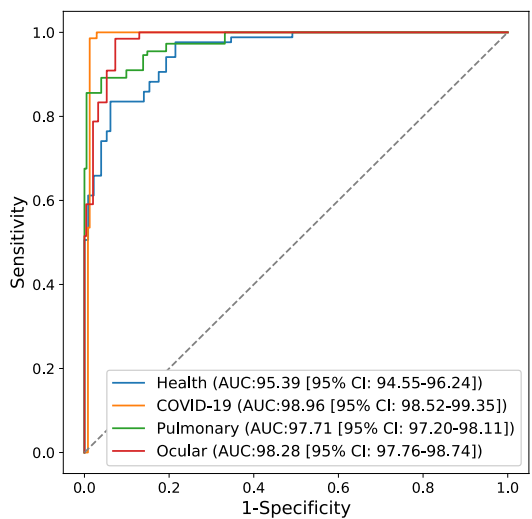

Figure B. The Screening Results of our method.

**Dataset**.

The data for the study comes from the following three resources,

- COVID-19 patients: all COVID-19 cases were acquired from May to July, 2020 by the respiratory physician of Department of Critical Care, Shanghai Public Health Clinical Center, Fudan University, Shanghai, China. All patients were diagnosed according to the diagnostic criteria of the National Health Commission of China and confirmed by RT-PCR detection of viral nucleic acids. COVID-19 patients were admitted to a negative pressure and specially ventilated room. The number of health care workers involved in the patient's care should be limited. All health care workers should use personal protective equipment appropriate for standard, contact, and airborne precautions (i.e., hand hygiene, gown, gloves, and N-95 respirator) in addition to eye protection while caring for these patients. The doctor at the bedside uses common smartphone to take photos of the eyes of COVID patients. After each patient takes pictures, the smartphone are disinfected to prevent cross infection.

- Finally, the photos as well as the patient's information are collected. Meanwhile, we obtained the epidemiological, medical history, clinical characteristics, laboratory tests and treatment history of the enrolled patients from electronic medical records and nursing records. Especially, all patients were given chest X-rays or CT. Some of the data needs to be supplemented and confirmed, and we obtain the data through direct communication with the doctor at the bedside. The details of demographics, basic characteristics, clinical characteristics and outcomes of the collected COVID-19 patients are summaries in Table. The images are captured by two smartphone

- The patients with other pulmonary diseases: all cases were obtained from May to July 2020 by the physician of Shanghai Public Health Clinical Center, Fudan University, Shanghai, China. These patients, diagnosed as pulmonary fungal infection, bronchopneumonia, chronic obstructive pulmonary and lung cancer, etc., have clear examination reports.

- The ocular patients: all cases were collected from June to July 2020 by the physician of Shanghai Public Health Clinical Center, Fudan University, Shanghai, China. These patients also have clear examination reports, and they are diagnosed as trachoma, pinkeye, conjunctivitis, glaucoma, cataract and keratitis, etc.

- Healthy people: the healthy cases came from the volunteers who had no disease or symptoms. In the process of volunteer recruitment, we balanced the distribution of age and gender. All volunteers were taken images according to our data-collection guidelines, so there is no uniform device category, which also simulates a real application scenario.

All participants were provided with written informed consent at the time of recruitment. And this study was approved by the Ethics Committee of Shanghai public health clinic center of Fudan University.

### Summary of the R&D

According to the official confirmed COVID-19 clinical manifestation, the current definitive method for COVID-19 diagnostic test includes nucleic acids, antibodies, serum, virus gene sequencing, CT imaging, history of epidemiology, clinical manifestation, etc.. Unfortunately, such tools are too expensive to be widely utilized in regions with scarce medical resource and yet inefficient to asymptomatic COVID-19 patients, who have been accounted for about 40% of the world's COVID-19 patients. Thus, it is of great value for a new criterion and method of COVID-19 risk screening and diagnosis.

In this study, we reviewed the signs and symptoms of hospitalized COVID-19 patients with a special focus on ocular manifestations. There are few published studies on eye involvement in COVID-19. Some studies have shown that the COVID-19 patients usually accompanied by ocular manifesting, history of epidemiology, clinical manifestation, etc.. Inspired by these, we propose a truly rapid COVID-19 risk screening and diagnosis model with deep learning method based on eye-region images, captured by normal CCD or CMOS camera and cellphone.

In this study, Our model can successfully classify COVID-19 patients from healthy persons, pulmonary patients except for COVID-19 (e.g., pulmonary fungal infection, bronchopneumonia, chronic obstructive pulmonary disease, and lung cancer) , and ocular patients. The experimental results reveal that patients with COVID-19 have different ocular features from others, which can be used to distinguish them from the public. The convenient method of eye-region image diagnosis can help disease control researchers to fully understand the prevalence and pathogenicity of the virus in different ages, time, region, climate, environment, occupation, and population with basic diseases, and guide effective prevention and control measures against COVID-19.

### Methodology

To evaluate the study, we conduct five-fold cross-validation experiments on a dataset：healthy people, COVID-19 patients, pulmonary patients, ocular patients. As for the COVID-19 category, the average sensitivity target is no less than 80%, which could demonstrate that the efficacy of our method to distinguish COVID-19 patients.

Technological realization includes ocular feature extraction and classification, construction of phenotype encoding criteria, annotation of ocular feature, image detection model training, double-blind tests verification, collection of users' ocular images, automatic matching and feedback of users' ocular images, few(zero)-shot ocular phenotype feature learning, the methodology of high-dimensional statistical modeling as well as pathology.

In the high-dimensional statistical inference part, the dimension of ocular feature data is far more than the sample number of COVID patients, which raises the problems of high-dimensionality, heterogeneity, noise accumulation, false correction, and endogeneity. We make analysis from these factors and separate effective information from noise in those high-dimensional data. And our model outputs the ocular surface characteristics from COVID patients. The solution is supposed to be sparse and of a high confidence interval, which provides the interpretability to the ocular phenotype of COVID as well as achieving certain accuracies.

The framework has two major components: an Image Preprocessing method to detect and crop the eye area from the input image, and a DL-based Classification Network to extract discriminative features and recognize COVID-19 patients based on the eye-region data (Fig. 1). To develop and validate our model, we collect a dataset with COVID-19 patients, healthy people, patients with other pulmonary diseases and ocular patients, and the data of each patient all come from cellphones, shown in Fig. 2. We apply five-fold cross-validation, all data is randomly partitioned into five-equal sized subsets, and there is no identity overlap among the subsets. Of the five subsets, one subset is utilized as the testing data, and the remaining four subsets are used for training the model. The cross-validation process is then repeated 325 times, with each of the subset used exactly once as the testing data. Then the five results can be averaged to produce a single estimation.

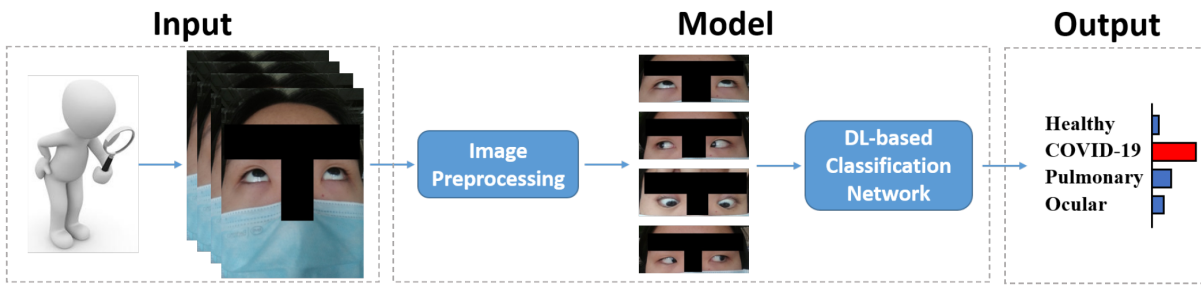

**Figure 1.** Illustration of the modeling framework.

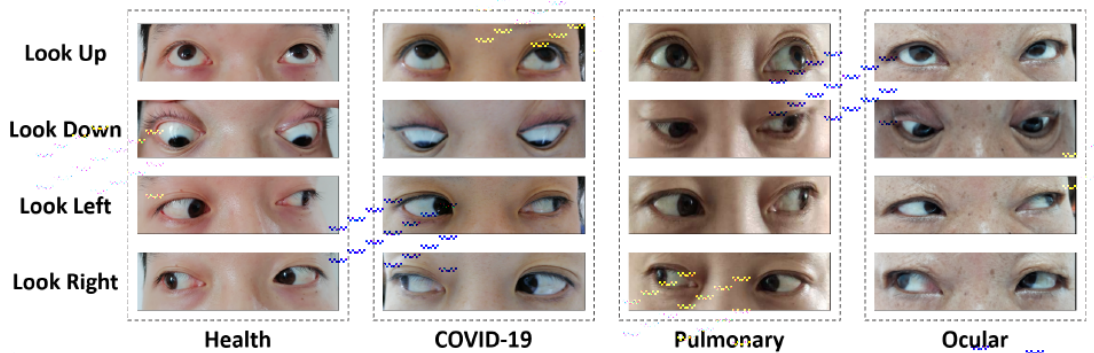

Figure 2. The examples of each patient.

## Statistical Analysis

For the statistical analysis, we apply lots of metrics to thoroughly evaluate the performance of the model. Following standard protocol, we totally utilize eight metrics, i.e., the sensitivity, specificity, accuracy, false positive error, false negative error, false disease prediction error, receiver operating characteristic curves, and area under the curves. Concretely, 'sensitivity', known as the true positive rate (TPR), indicates the percentage of positive patients with correct discrimination. Referred to the true negative rate (TNR), 'specificity' represents the percentage of negative persons who are correctly classified. 'Accuracy' is the percentage of the number of true positive (TP) and true negative (TN) subjects. The 'false positive/negative error' (FPE/FNE) measures the percentage of negative/positive persons who are misclassified as positive/negative. 'False disease prediction error (FDPE) calculates the percentage of positive persons whose disease types (i.e., COVID-19 and Others) are predicted incorrectly. Receiver operating characteristic curves (ROC) and area under curves (AUC) is used to show the performance of the classifier.

**Conclusion**.

COVID-19 is an infectious disease caused by SARS-CoV-2. The patients usually have fever, cough, fatigue, shortness of breath, and loss of smell and taste. Nowadays, some researches also find that the Olfactory and Gustatory Dysfunction is one of the common symptom of the COVID-19. There are also exists some works to verify that SARS-CoV-2 might be transmitted through the eye, the testing results for SARS-CoV-2 from conjunctival swabs can be positive and patients may have the symptom of bilateral acute conjunctivitis. Therefore, considering that such symptoms may be ignored during the diagnosis, we pay attention to using eye-region features to help the risk assessment.

Consider that eye exam technology has been used to screen for a variety of diseases, such as diabetes and kidney disease. In this paper, we proposed a deep learning model for rapidly risk screening COVID-19 with eye-region images. Different from previous studies, which utilize RT-PCR or CT imaging, the input of our system is the face image or binocular image captured by common CCD cameras. Combining with the development of deep learning, it enables the real-time COVID-19 screening from two aspects, sample acquisition and testing.

We are conducting large-scale experiments to further validate the effectiveness and efficiency of our algorithm as the tool of new screening method. On the other hand, due to the privacy policy and the difficulty of data collection in non-Asian patients, we will further investigate and extend the capability of our algorithm in few-shot learning settings. We believe that this study can be inspiring and helpful for encouraging more researches in this direction, and provide effective and rapid assist for clinical risk screening, especially during outbreaks.

To conduct further research, we also implement an online screening trial platform which implements our algorithms, and help fast screening COVID-19. For more information, please contact Dr. Yanwei Fu yanweifu@fudan.edu.cn. We hope that our algorithm and system should help the rapid screening of COVID-19 globally.

**Table 1.** Demographics, basic characteristics, clinical characteristics, and outcomes of 104 patients with COVID-19.

|  | Patients (n=104) |
|---|---|
| **Age (years)** | |
| Mean | 34 |
| Range | 5 ~ 66 |
| **Sex** | |
| Female | 34 (32.69%) |
| Male | 70 (67.31%) |
| **Chronic medical illness** | |
| Hypertension | 7 (6.73%) |
| Diabetes | 3 (2.88%) |
| Asthma | 8 (7.69%) |
| Malignant tumour | 1 (0.96%) |
| Others | 8 (7.69%) |
| **Signs and symptoms at admission** | |
| Yes | 78 (75.00%) |
| No | 26 (25.00%) |
| **Chest x-ray and CT findings** | |
| Normal | 14 (13.46%) |
| Unilateral pneumonia | 12 (10.91%) |
| Bilateral pneumonia | 84 (76.36%) |
| Multiple mottling and ground-glass opacity | 52 (50.00%) |
| **Clinical classification** | |
| Asymptomatic | 4 (3.85%) |
| Light | 15 (14.42%) |
| Recurrence of SARS-CoV-2 nucleic acid positive | 1 (0.962%) |
| Ordinary | 82 (78.85%) |
| Critical | 2 (1.92%) |
| **Clinical outcome** | |
| Discharged | 104 (100%) |
| Died | 0 (0%) |

## Sample Collection

For scientific research, we need some eye region data. We promise that the data will not be used for commercial, only for scientific research. The specific requirements are as follows:

1. The eyes in the image need to be clear, a total of five required angles are shown on the left.

2. When taking photos, please do not wear cosmetic products such as contact lenses.

3. When taking photos, please do not use beauty camera mode, or post-production beauty filters, we need the original images.

1. 下视 look down
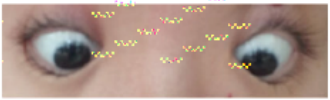

2. 平视 look horizontally
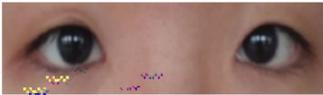

4. 右视 look right
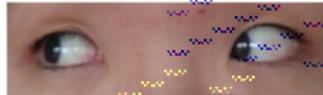

3. 左视 look left
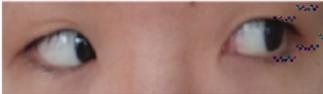

5. 上视 look up
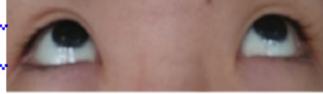

**1. The number of tested COVID-19 patients is 300.**

1) Among COVID-19 patients, the ratio of "asymptomatic infection (nucleic acid positive): mild/common: severe/critical: turn negative: re-positive" is recommended to be 3:3:3:1:1:

2) White (including Hispanic): Black: Asian (including Indian and Middle Eastern), the recommended ratio is 2:2:1

**2. The number of negative sample control groups (negative samples) is 300**

Patients with non-neo-COVID-19 lung diseases: patients with eye diseases: healthy people, the recommended ratio is 1:1:1.

If the work cost is too high, the negative sample control group can not be provided with the patient's name, medical history, all faces, and other data related to personal privacy and biosafety information. If necessary, the experimenter can wear a mask to take pictures.

### 3. Sampling equipment and environmental requirements

1) For the above sampling, the same model of mobile phone or shooting equipment must be used to prevent the sampling data domain from being interfered with by the equipment.

2) If the conditions are not available, it is necessary to use the same model of mobile phone to collect COVID-19, non-COVID pneumonia, eye diseases, and healthy people at the same time, to maximize the elimination of device data domain interference.

3) The same shooting parameters must be used when shooting, and the beauty, soft light, and other shooting filters must not be used.

4) The image resolution of the eyes is at least: 1900x500 96dpi

5) The shooting environment should be well-lit and bright. It should not shoot in front of dark and red backgrounds. A white background is best.

### 4. Sampling compliance

a) The subject or legal representative has signed an informed consent form

b) Age ≥18 and ≤75 years old (under 18 years old, the guardian shall sign the informed consent form)

c) Meet the diagnostic criteria for COVID-19 infection

**Ethics committee approval**

All participants were provided with written informed consent at the time of recruitment. And this study was approved by the Ethics Committee of Shanghai public health clinic center of Fudan University. The study was in accordance with the Declaration of Helsinki 1964 and its successive amendments.

**Conflict of interest statements**

All other authors declare no competing interests.

**Acknowledgments**

We would like to thank Dr. Zhilin Yang, Dr Yunbin Zhang, and Mr. Yu Liu for their kind assistance with this research project.